\documentclass[lefteqn,showpacs,preprint]{revtex4}

\usepackage{graphicx}
\usepackage{dcolumn}
\usepackage{bm}

\begin{document}
\draft
\title{Zero-refraction in natural materials and the mechanism of metal superlens}
\author{Yun-Song Zhou$^{1,\dag}$, Huai-Yu Wang$^2$, Hai Wang$^1$}
\date{\today }
\draft
\address{$^1$ Center of Theoretical Physics, Department of Physics, Capital Normal University. Beijing
100048, China}
\address{$^2$ Department of Physics, Tsinghua
University, Beijing 100084, China}

\begin{abstract}
We found that a single negative material has a character of zero-refraction in near field, and point out that the mechanism of a metal superlens to image with the resolution exceeding the diffraction limitation is different from that of a perfect lens made of a double negative material. The principle of metal superlens is disclosed. Our numerical results based on the zero-refraction character coincide with the experimental results well. This work brings new understanding about the single negative materials and will lead the applications of metal superlens in right way.
\end{abstract}


\maketitle

\newpage

All electromagnetic materials can be sorted into four classes according to their permittivity $\epsilon$ and permeability $\mu$ , i. e. the double positive (DPS, $\epsilon >0, \mu >0$) , electric negative (ENG, $\epsilon <0, \mu >0$ ), double negative (DNG, $\epsilon <0, \mu <0$  )[1], and magnetic negative (MNG,  $\epsilon >0, \mu <0$) materials, just as shown in Fig. 1. It has been realized that the first (DPS) and third (DNG) quadrants correspond to positive and negative refractions[1-13], respectively. In the second (ENG) and fourth (MNG) quadrants, media are opaque. On the $\epsilon$ and $\mu$  axes, as $n=\sqrt{\epsilon}\sqrt{\mu}=0$  , the corresponding materials show the zero-refraction character. The zero-refraction material is expected to have great potential in applications, but it seems that the zero-refraction materials were never found in nature. However, the above conclusion is valid in far field. The near field character of a single negative (SNG) materials, which is either an  ENG or a MNG material, has not been clarified. In 2000, Pendry [14] predicted that the superlens made of SNG slab could replace the perfect lens made of DNG slab to challenge the diffraction limitation [14,15]. Following the ideal, Zhang et al. experimentally obtained the sub-diffraction-limited image with one-sixth of the illumination wavelength by using a silver film as the superlens [16]. After that, based on the superlens properties, the magnifying optical superlens consisting of a curved periodic stack of Ag and dielectric was realized [17,18]. As Ag film is an ENG material, easer to be fabricated and less expensive than DNG materials, it is called the "poor man's superlens" [19]. These results paved a way for the applications of SNG materials in nanoscale optical imaging and ultrasmall optoelectronic devices. For instance, the optical lithography technology of micro electric circuit will be preceded to nanometer size to raise the storage density[20-22].
\begin{figure}
\begin{center}
\includegraphics*[width=8cm]{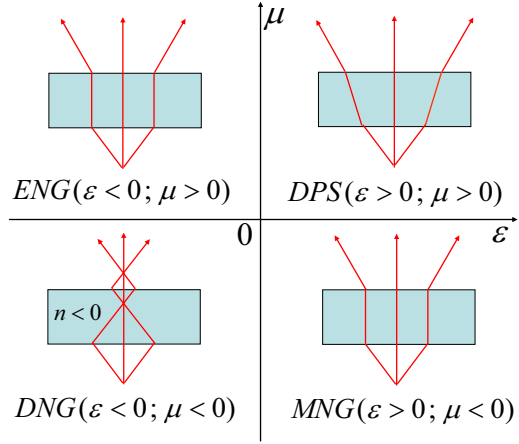}

\end{center}
\caption{ Four kinds of refractions in near field according to the values of permittivity $\epsilon $ and permeability $\mu$ . } \label{TERS_Setup}
\end{figure}

Nevertheless, the refraction in the second and fourth quadrants in the near field has to be elucidated. Although a DNG slab can focus light from a point source by negative refraction, just as displayed in the third quadrant of Fig. 1, one cannot assume the case as a matter of course in ENG and MNG materials, because the refraction characters in the second and fourth quadrants are not clear. We find that the SNG materials correspond to the zero-refraction. This discovery is applied to disclose the principle of metal superlens, which is quite different from that of a perfect lens proposed by Pendry [14,15]. The resolution of the metal superlens used by Zhang et al. [16] is numerically calculated based on the zero-refraction, and their experimental results are well explained.

First of all, let us consider an isolated ideal SNG slab with dielectric constant  $\epsilon <0 $ and thickness $d$. A Cartesian coordinate is set with the $z$ axis perpendicular to the slab and the origin is at the incident surface. In an S-polarized incident light, the electric fields is written as
$$
\textbf{E}_{1S}^{i}(x,z)=\exp(ik_{x}x+ik_{1z}z-\omega t)\hat{y}. \eqno   (1)
$$
Here the incident amplitude is assumed to be 1. Then at the two sides of incident surface the reflected and transmitted waves can be written as
$$
\textbf{E}_{1S}^{r}(x,z)=r_{12}\exp(ik_{x}x-ik_{1z}z-\omega t)\hat{y} \eqno   (2)
$$
and
$$
\textbf{E}_{2S}^{t}(x,z)=t_{12}\exp(ik_{x}x-|k_{2z}|z-\omega t)\hat{y}, \eqno   (3)
$$
respectively, where
$$
|k_{2z}|=\sqrt{(\omega/c)^{2}|\epsilon_{2}|+k_{x}^{2}} \eqno (4)
$$
and
$$
k_{1z}=\sqrt{(\omega/c)^{2}-k_{x}^{2}}. \eqno (5)
$$
The coefficients $r_{12}$  and $t_{12}$  are the amplitude reflection and transmission coefficients, respectively, at the incident surface. After the transmission at the exit surface, the electric field reads
$$
\textbf{E}_{3S}^{t}=t_{12}t_{23}\exp(-|k_{2z}|d)\exp[ik_{x}x+ik_{1z}(z-d)-\omega t]\hat{y}. \eqno   (6)
$$

It is seen from Eq. (3) that the wave in the slab decays while keeping the phase along $z$ direction unaltered.

However, the above results are obtained for a plane wave. When considering the imaging process, one must know the optical path of a thin beam. Suppose that a narrow beam with width $b$  and incident angle $\theta$   impinges on the incident surface. The incident beam is described by
$$
\textbf{E}_{1S}^{i}(x,z)=F(x,z)\exp(ik_{x}x+ik_{1z}z-\omega t)\hat{y}, \eqno   (7)
$$
where $F(x,z)$ is the rectangle function. On the incident surface, $z=0$, $F(x,z=0)$ it is expressed as
$$
F(x)=\{\begin{array} {ccc}
1/\sqrt{b/\cos\theta}  & , x\leq |b/(2\cos\theta)|\\
0  & ,x > |b/(2\cos\theta)|%
\end{array}%
.\eqno (8)
$$
Therefore the electric field distribution at the incident surface (at time $t=0$) becomes
$$
\textbf{E}_{1S}^{i}(x,z)=F(x)\exp(ik_{x}x)\hat{y}. \eqno   (9)
$$
This electric field can be expanded by plane waves, and each plane wave component corresponds to one transmission field described by Eq. (3). The total field in the slab should be the integral of these transmission fields, so we write

$$
\textbf{E}_{2S}^{t}(x,z)=\frac{1}{2\pi}\int_{-\infty}^{\infty} dk^{\prime}_{x}\int_{-b/(2\cos\theta)}^{b/(2\cos\theta)}dx^{\prime}t_{12}\exp[i(k_{x}-k_{x}^{\prime})x^{\prime}]\exp(ik_{x}^{\prime}x-|k_{2z}^{\prime}|z)\hat{y} , \eqno   (10)
$$
where $k_{x}^{\prime}$ and $k_{2z}^{\prime}$  meet Eq. (4), $|k_{2z}^{\prime}|=\sqrt{(\omega/c)^{2}|\epsilon_{2}|+k_{x}^{2}} $ .

Now we give numerical examples to show how the transmitted thin beam propagates in the slab from the incident to output surfaces. In general, $t_{12}<1$ , but for present purpose, we may take $t_{12}=1$. The metal is Ag and the slab thickness is set as $0.035 \mu m$. The incident wavelength is $\lambda_{0}=0.365\mu m$ , and accordingly the permittivity of Ag is $\epsilon=-2.4012+i0.2488$[15] . The beam width is $b=10\lambda_{0}=3.65\mu m$ . In this case, the material is nonideal, but the imaginary part can be neglected since it is less than the real part by one order of magnitude, i.e., the metal film is approximated by an ideal ENG material. The calculation results for the cases of $\theta=0$  and $\theta=30^{\circ}$  are displayed in Fig. 2. Figure 2(a) shows the case of normal incident $\theta=0$ . The profile of the electric field at the incident surface is exactly rectangular. When the wave reaches the exit surface, the profile is still a rectangle with the width unchanged, although the intensity is weakened and the edges become blurry faintly. As for the case of inclined incidence  $\theta=30^{\circ}$ (see Fig. 2(b)), the electric field shows the oscillation distribution on both incident and output surfaces, with less amplitudes on the latter. Thus we achieve the knowledge that, the profile of the electric field is copied from the incident plane to exit plane without any shift along $x$ direction, but only the intensity decreases. In other words, when a thin light beam strikes on the surface of an ENG slab, it will transmit along the direction perpendicular to the surface. The refraction angle is always zero independent of incident angle.
\begin{figure}
\begin{center}
\includegraphics*[width=12cm]{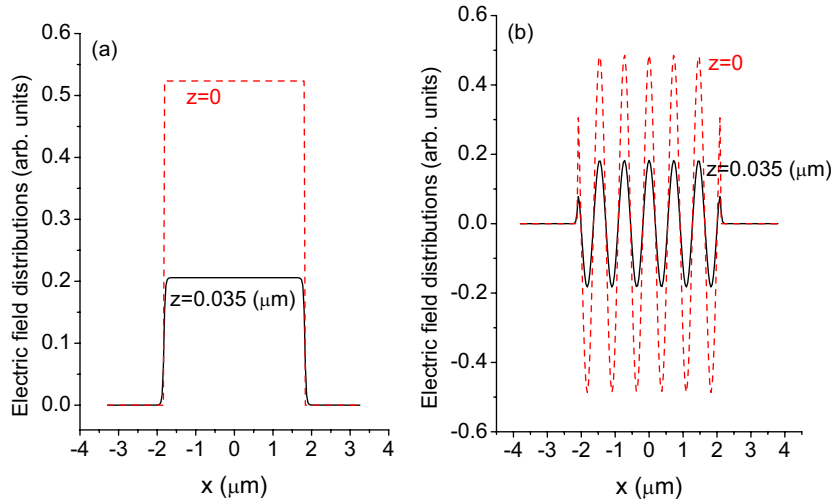}

\end{center}
\caption{ The profiles of the electric field at incident surface (dashed curves) and at output surface (solid curves) of a thin Ag slab when the incident angle is (a) $\theta=0$. and (b) $\theta=30^{\circ}$ } \label{TERS_Setup}
\end{figure}

It can be verified in the same way that the zero refraction will also occur when light goes through the surface of an MNG material. We outline the general conclusion concerning refraction with materials as follows: the refractive angle is positive in a double positive (DPS) material (positive refraction), negative in a DNG material (negative refraction) and zero in a SNG material, as depicted in Fig. 1. This conclusion is valid for both S- and P-polarized waves.

As an application example of our discovery, we propose the superlens theory based on zero-refraction character. The installation utilized in experiment [16] is drawn in Fig.3. An Ag slab is sandwiched by two dielectrics with the permittivitys $\epsilon_{1}$  and $\epsilon_{3}$ , respectively. For convenience, we label the regions of $\epsilon_{1}$  ,$\epsilon_{2}$ , and $\epsilon_{3}$ by characters I, II, and III, respectively. A Cartesian coordinate is set with the z axis perpendicular to the metal film. The interfaces I/II and II/III are the incident and output surfaces of the metal slab.

\begin{figure}
\begin{center}
\includegraphics*[width=8cm]{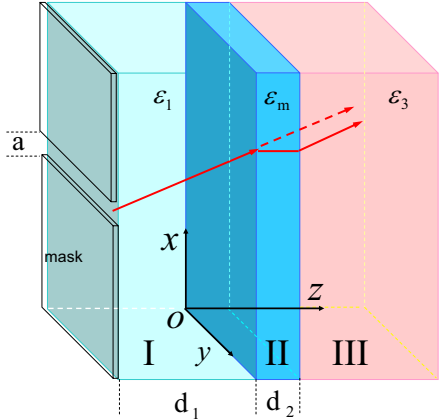}

\end{center}
\caption{ The sketch of the superlens set in the photolithography experiment.} \label{TERS_Setup}
\end{figure}

Before doing numerical calculation, let us consider the working principle of the superlens based on the zero-refraction theory. The red arrow in Fig. 3 indicates a light line from the mask slit, which is zero-refracted by the metal slab and then arrives at the surface of $\epsilon_{3}$  material. Without the metal slab, the light would go along the dashed arrow and reach the surface of $\epsilon_{3}$  material at the point farther from the mask slit. Obviously the metal slab can beam the light to form a narrower image of the mask slit, which is termed as the beaming effect.

Another effect is that for the radiation wave, the transmissivity will monotonously decrease with the increase of the tangential component of the wave vector. This character acts as an angular filter to weaken the light with a larger incident angle so as to concentrate the energy at the central region. This effect was in fact first obtained by Pendry [14] when he deduced the total amplitude transmissivity of a DNG slab, and later discussed in Ref. [16].

Having analyzed the physical mechanism of superlens, we give the numerical calculation of the image profile. The model coincident to the photolithography experiment in Ref. [16] is sketched in Fig. 3. The electric field distributions of S and P waves on image material surface (interface II/III) are
$$
\textbf{E}_{S}(x,z)=\int_{-a/2}^{a/2}H_{0}[(2\pi\sqrt{\epsilon_{1}}/\lambda_{0})\sqrt{(x-x_{1})^{2}+d_{1}^{2}}]T_{S}(x-x_{1})dx_{1}
 , \eqno   (11)
$$
and
$$
\textbf{E}_{P}(x,z)=\int_{-a/2}^{a/2}H_{0}[(2\pi\sqrt{\epsilon_{1}}/\lambda_{0})\sqrt{(x-x_{1})^{2}+d_{1}^{2}}]T_{P}(x-x_{1})\sqrt{\epsilon_{1}/\epsilon_{3}}dx_{1}
 , \eqno   (12)
$$
respectively. Here $H_{0}$  is the zero-order Hankel function, and $T_{S}$ , $T_{P}$  the amplitude transmissivities of the metal slab for S and P-waves, respectively. Similar to the way used in Ref. [13], we obtain the transmissivities of the metal slab sandwiched by dielectrics I and III:
$$
T_{S,P}=\frac{t_{12}^{S,P}t_{23}^{S,P}\exp(ik_{2z}d_{2})}{1-r_{21}^{S,P}r_{23}^{S,P}\exp(i2k_{2z}d_{2})}.\eqno (13)
$$
The interface transmissivities and reflectivities are in turn expressed as
$$
r_{i,j}^{S}=\frac{k_{iz}-k_{jz}}{k_{iz}+k_{jk}}.\eqno (14)
$$
$$
t_{i,j}^{S}=\frac{2k_{iz}}{k_{iz}+k_{jk}}.\eqno (15)
$$
$$
r_{i,j}^{P}=\frac{k_{iz}/\epsilon_{i}-k_{jz}/\epsilon_{j}}{k_{iz}/\epsilon_{i}+k_{jk}/\epsilon_{j}}.\eqno (16)
$$
$$
t_{i,j}^{P}=\frac{2k_{iz}/\epsilon_{i}}{k_{iz}/\epsilon_{i}+k_{jk}/\epsilon_{j}}.\eqno (17)
$$
where $i=1,2$ and $j=i+1$. In calculation, $k_{iz}$  is related to $k_{ix}=\sqrt{\epsilon_{i}}(\omega/c)\sin\arctan[(x-x_{i}/d_{2})]$ , by the common electromagnetism relationship just as, for instance, Eq. (4). For a linear image material, its notch depth is proportional to the striking light energy, and the energy distribution is proportional to the square of amplitude of total electric field: $|E|^{2}=|E_{S}|^{2}+|E_{P}|^{2}$.

The parameters are according to the experiments [16] taken as follows: the mask slit width is $a=40nm$; the incident
wavelength $\lambda_{0}=365nm$; $\epsilon_{1}=2.3013+i0.0014$ for PMMA; $\epsilon_{2}=-2.4012+i0.2488$ for silver
superlens; $\epsilon_{3}=1.5170+i0.00046$ for image material. On the image material, the exposed notch with the
half-maximum width (HMW) of $w_{1}=89nm$ was observed. When the superlens was replaced by a PMMA with thickness
$35nm$, the HMW became $w_{2}=321nm$ . This shows that the superlens has narrowed the HMW from $321$ to $89nm$.
It is appropriate to define a ratio $\eta=w_{2}/w_{1}$ , named as narrowing factor, to describe the narrowing effect.
In the experiment, this ration is  $\eta_{e}=321/89\approx3.6$.

In Fig. 4 displayed are our calculated energy distributions at the right side of II/III interface. The distributions
are proportional to the depths of the exposure notch. The solid and dashed curves correspond to the result with and
without the superlens, respectively. The corresponding HMWs are $72$ and $246nm$, respectively, both being narrower
than those in experiment by about $20$ percent. We believe that the experimental wider HMWs resulted from the factors such as the roughness of the interface between the superlens and dielectrics, the mask slit edges and so on. Nevertheless, our calculated narrowing factor is $\eta_{t}\approx3.4$, quite close to the experimental result.
\begin{figure}
\begin{center}
\includegraphics*[width=12cm]{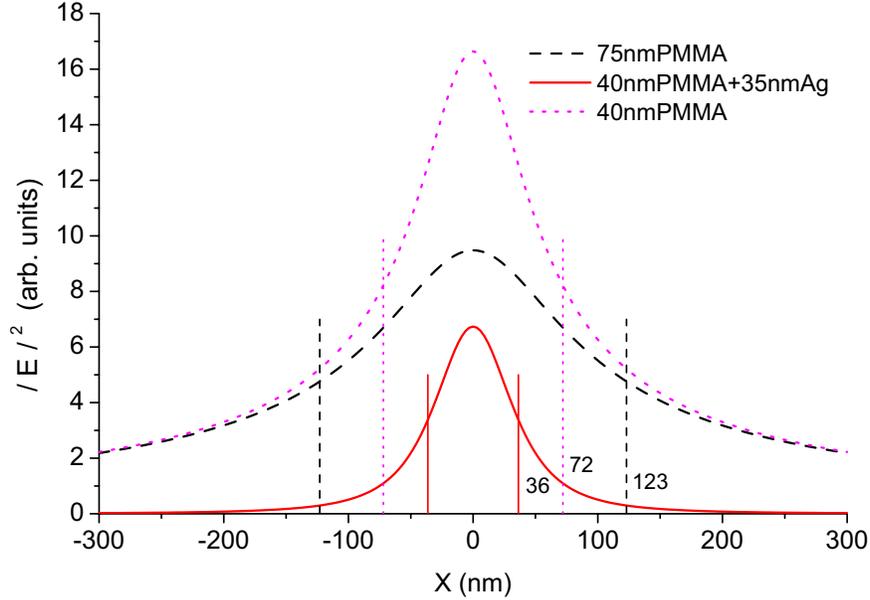}

\end{center}
\caption{ The energy distributions on the surface of the image material (right side of interface II/III in Fig. 3)} \label{TERS_Setup}
\end{figure}

If the superlens is removed from the experimental set while the PMMA thickness being kept unchanged, the image material will separated by the $40nm$ PMMA from the mask, the corresponding theoretical distribution of the exposure depth is expressed by the dotted line in Fig. 4. Its HMW, being $144 nm$, is narrower and the peak is higher than the dashed curve. Since the image material is closer to the mask slit when the metal slab is absent, the exposure distant becomes shorter, so that the light beam is more concentrated. When the superlens is inserted between the image material and PMMA, the exposure distance will be extended. The extension does not cause the light diffusion due to the zero refraction. The filter effect whittles the light with lager incident angle, which causes the exposure notch to be narrowed to the solid curve in Fig.4.

In summary this paper makes the following contributions: (i) in the near field, the ENG and MNG materials are of the zero-refraction character. This finding filled the cognizing gap about the natural materials sorted by their electromagnetic properties. The zero-refraction was realized only in metamaterial, such as photonic crystals, but never in natural material before; (ii) the principle of superlens made of an ENG or MNG material is disclosed, which enhances the image resolution by beaming and filtering effects, instead of the focusing effect in perfect lens; (iii) a method is proposed to calculate the resolution of superlens, and our numerical results are in good agreement with experiment.

\vskip8pt \textbf{Acknowledgements}

\vskip5pt This work is supported by the 973 Program of China (Grant
No.2011CB301801) and the National Natural Science Foundation of
China (Grant No. 10874124), and (Grant No. 11074145).





\[
\]

\end{document}